\documentclass[twocolumn,superscriptaddress,showpacs,amsmath,amssymb]{revtex4} 
\usepackage{graphicx}% Include figure files
\usepackage{dcolumn}% Align table columns on decimal point
\usepackage{bm}% bold math

\begin{document}

\title{Tapering Enhanced Stimulated Superradiant Amplification}

\author{J. Duris}
\affiliation{Department of Physics and Astronomy, UCLA, Los Angeles, CA, 90095, USA}
\author{A. Murokh}
\affiliation{Radiabeam Technologies, Santa Monica, CA 90404, USA}
\author{P. Musumeci}
\affiliation{Department of Physics and Astronomy, UCLA, Los Angeles, CA, 90095, USA}

\date{\today}

\begin{abstract}
High conversion efficiency between electrical and optical power is highly desirable both for high peak and high average power radiation sources. In this paper we discuss a new mechanism based on stimulated superradiant emission in a strongly tapered undulator whereby an optimal undulator tapering is calculated by dynamically matching the resonant energy variation to the ponderomotive decelerating gradient. The method has the potential to allow the extraction of a large fraction ($\sim$50\%) of power from a relativistic electron beam and convert it into coherent narrow-band tunable radiation, and shows a clear path to very high power radiation sources.

\end{abstract}

%41.60.Ap Synchrotron radiation
%41.60.Cr Free-electron lasers
%41.75.Jv Laser-driven acceleration
%41.75.Lx Other advanced accelerator concepts

\maketitle

%\section{Introduction}
Among coherent radiation sources, free-electron lasers (FEL) carry unique advantages such as wavelength tunability and access to the short wavelength region of the electromagnetic spectrum. FELs are not limited by thermal or non-radiative loss mechanisms characteristic of atomic lasers based on solid state and gas phase gain mediums. Nevertheless, saturation effects limit the conversion efficiency to levels comparable with the Pierce parameter $\rho$ which is typically $<0.1\%$ \cite{BPN}. FEL undulator tapering \cite{KMR} has been shown to allow much larger efficiencies. At very long wavelengths (35~GHz) where it is possible to use a waveguide to maintain an intense radiation field on axis, up to 35\% conversion efficiency has been demonstrated \cite{Scharlemann:ELF}. At shorter wavelengths \cite{Wu:prstab,CEmma:TWLCLS}, the reduction of gain guiding and the onset of spectral sidebands have limited the effectiveness of tapering \cite{Sprangle:FELsidebands}. For example at the LCLS, the power extraction has remained below the percent level limiting the amount of energy in the pulse to a few mJ. Higher conversion efficiencies could lead to unprecedented intensity X-ray pulses with over 10$^{13}$ photons per pulse providing sufficient signal-to-noise to enable the long sought goal of single molecule imaging \cite{4thgeneration:lcls}. In the visible and UV spectral ranges, large electrical to optical conversion efficiencies are also very attractive for the development of high average power (10-100 kW-class) lasers especially when considering that superconducting radio-frequency linacs can create relativistic electron beams with very high wall-plug efficiencies and MW average power.

In order to increase the electro-optical conversion efficiency, we note that the burgeoning field of laser accelerators is making extreme progress on the opposite problem---that is, optical to electrical power conversion. Among the various schemes for laser acceleration, the inverse free-electron laser (IFEL) accelerator is a far-field vacuum-based scheme which uses an undulator magnet to couple a transversely polarized radiation field to the longitudinal motion of the electrons \cite{Palmer:IFEL}. The lack of nearby boundaries or a medium (gas, plasma) to couple the light to the electrons implies very little irreversible losses and in principle enables very high energy transfer efficiencies. Simulations show that an IFEL could be optimized to transfer 70\% of optical power to a relativistic electron beam \cite{GeVIFEL}. Recent experimental results demonstrated energy doubling of a 52~MeV beam with $\sim$100 MeV/m average accelerating gradients and capture of up to 30\% of the injected electron beam \cite{Rubipaper} using a strongly tapered undulator in a helical geometry IFEL interaction. Reversing the process to decelerate the beam by the same mechanism, it would be in principle possible to transfer back to the drive laser half of the electron energy---effectively, extracting 50\% of the electron beam power and converting it into coherent radiation \cite{Nocibur}.

Based on this idea, we investigate in this paper a novel scheme for efficient generation of radiation whereby a high intensity seed laser pulse and a relativistic electron beam copropagate in a tapered undulator and the IFEL interaction is used to decelerate the beam. The scheme relies on the coherent emission of a prebunched beam going through an undulator in the presence of an intense driving field (i.e. stimulated superradiance emission \cite{Gover:Superrad}). Very strong tapering of the undulator is the other key ingredient to enable high conversion efficiencies and support large deceleration gradients and electron energy losses.

%\begin{figure}
%\includegraphics[width=80 mm]{fig1-schematic.png}
\begin{figure*}[t]
\includegraphics*[width=175 mm]{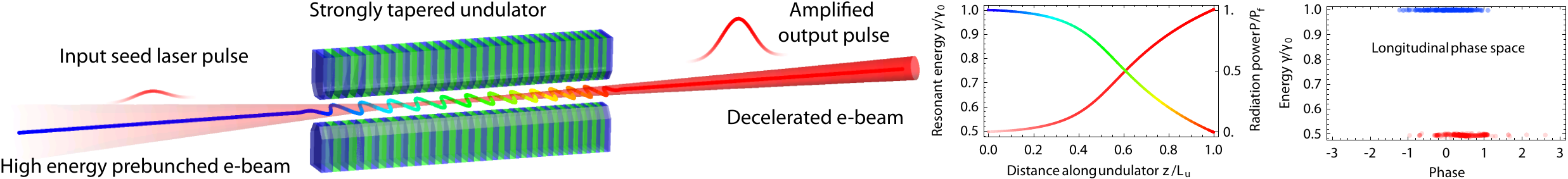}
\caption{An implementation of TESSA. The diagram shows a seed laser focused into the strongly tapered undulator along with a prebunched electron beam. As the e-beam energy---high energy being blue while low is red---decreases along the undulator, the laser power grows. Plots of normalized electron energy and laser power along the undulator and initial and final longitudinal phase spaces are shown. }
\label{Figure:Schematic}
\end{figure*}

This tapering-enhanced stimulated superradiant amplification (TESSA) can be viewed as essentially an IFEL run backwards (see Figure \ref{Figure:Schematic}). The drive laser field stimulating the electron emission can be obtained from an external low rep-rate seed laser, or in a spectral region where external sources are not available, from redirecting the saturated FEL radiation in a TESSA afterburner undulator \cite{TESSApatent}. In this case, in order to maximize the decelerating gradient, the FEL pulse can be refocused to reach peak intensities significantly larger than the FEL saturation level. In principle, it is also possible to obtain the seed pulse from the build-up in an oscillator cavity \cite{Avi:oscillator}.

%\section{Tapering equations}

The TESSA particle dynamics in longitudinal phase space are best understood in the formalism of a high-gradient IFEL decelerator. We begin the analysis with the 1D FEL/IFEL equations but limit the discussion to helical undulators as they offer at least twice the accelerating or decelerating gradient of planar ones.
\begin{subequations}\label{1dpureifel}
\begin{eqnarray}
\label{1dpureifela}\label{Equation:PondGrad}
\frac{d\gamma^2}{dz}&=& 2 k K_l K \sin \psi \\
\frac{d\psi}{dz} &=& k_w-k \frac{1+K^2}{2 \gamma^2}
\label{1dpureifelb}
\end{eqnarray}
\end{subequations}
Here, $\gamma$ is the particle energy in units of the rest energy $m_0 c^2$ while $K = e B_0 / m_0 c k$ and $K_l = e E_0 / m_0 c^2 k$ are respectively the undulator and laser field normalized vector potentials. The peak undulator magnetic field is $B_0$ while the peak laser electric field is $E_0$, $k = 2 \pi / \lambda$ and $k_w = 2 \pi / \lambda_w$ are the laser and undulator wavenumbers, and $\psi = (k_w + k) z - \omega t$ is the ponderomotive phase---that is, the phase of particles in the sinusoidal potential formed by the combined action of the laser and undulator fields on the electrons.

Resonant interaction requires a slowly varying ponderomotive phase $\psi_r$ throughout the undulator. This is achieved by requiring $d\psi/dz=0$ in Eqn. \ref{1dpureifelb} for a resonant particle, leading to the resonance condition $\gamma_r^2 = k (1+K^{2})/2 k_w$. The resonant phase is a key parameter for the system and determines many properties of the longitudinal dynamics such as the gradient and the extent of the stable trapping area in the phase space (bucket). In general $\psi_r$ is chosen around $-\pi/4$ to maximize extraction efficiency, finding the best compromise between deceleration and detrapping.

Taking the derivative of the resonant condition, a decelerating gradient can then be determined from the variation in the undulator parameters along $z$:
\begin{equation}
\label{Equation:ResEnergyGrad}
\frac{d\gamma_r^2}{dz} = \frac{1}{2\lambda}(1+K^2)\frac{d\lambda_w}{dz}+\frac{\lambda_w}{\lambda}K\frac{dK}{dz}
\end{equation}

For a given resonant phase $\psi_r$, there are different ways to optimize the tapering. For an undulator with constant period where $d\lambda_w/dz=0$, setting the resonant energy gradient in Eqn. \ref{Equation:ResEnergyGrad} equal to the ponderomotive gradient in Eqn. \ref{Equation:PondGrad} yields an equation for determining the optimum variation of $K$:
\begin{equation}
\label{Equation:KTaper}
\frac{dK}{dz} = 2 k_w K_l \sin \psi_{r}
\end{equation}

It is well known from IFEL accelerator experience that changing the undulator period brings further advantages, both in flexibility (one can keep a larger $K$ along the interaction), and in practical implementation since usually the magnetic field depends on the undulator period. Equating the right sides of Eqns. \ref{Equation:ResEnergyGrad} and \ref{Equation:PondGrad} for varying period yields
\begin{equation}\label{Equation:PeriodTaper}
\frac{d\lambda_w}{dz} = \frac{8 \pi K_l K \sin \psi_r}{1+K^2+\lambda_w dK^2/d\lambda_w}
\end{equation}
This tapering equation can be solved once an undulator builder equation relating $K$ to the undulator period $\lambda_w$ is given. While there are many undulator designs which may relate these parameters, one particular design is a helical permanent magnet Halbach undulator \cite{Halbach:Undulators} with an undulator builder equation given by $K \cong \frac{e \lambda_w}{2 \pi m c} 1.6 B_r e^{-\pi g/\lambda_w}$ where $g$ is the undulator gap and $B_r$ is the remnant field of the magnets. This specific undulator design is particularly useful and will be used in the examples below. Ultimately, the maximum energy extracted is limited by the feasibility of constructing an undulator with parameters matched to the resonance condition for the decelerated beam.

%\section{Low gain regime}

In order to reach a better understanding of the TESSA dynamics, we start by analyzing the low-gain regime where the radiation power does not significantly change along the undulator. Defining the efficiency as the relative change in energy for the beam $\eta = \gamma_f/\gamma_0-1$ and assuming $|\eta| \ll 1$ for the constant period tapering case, we obtain an estimate of $\eta \approx 2 \pi N_w K_l \sin \psi_r $. In practice to reach tens of percent efficiency, the number of periods in the undulator $N_w$ should be on the order of $K_l^{-1}$. Note that if the injected electron beam is not prebunched, the first section of the interaction can be designed with $\psi_r\approx0$ until full bunching occurs and the deceleration can start. Since the efficiency in the low-gain regime is independent of beam parameters, the output radiation power scales linearly with the input e-beam current. Considering diffraction, for a nearly constant undulator $K$, efficiency is maximized when a TEM$_{00}$ Gaussian seed laser is focused with a Rayleigh range of $z_r \approx 0.15 L_w$ to a waist at the undulator midpoint \cite{GeVIFEL}.

%\section{High gain regime & GITS}

Whenever the stimulated superradiant emission becomes the dominant contribution to the total laser field driving the interaction, the undulator can be tapered more aggressively in order to take advantage of the additional ponderomotive drive and maximize the deceleration gradient. In this case, the optimum tapering (and the conversion efficiency) depends on the injected e-beam current since a higher beam power generates more radiation which allows larger decelerating gradients and higher energy transfer rates. Thus in the TESSA high gain regime, the output radiation power grows faster than linearly with respect to input e-beam current.

The main difference in calculating the tapering is that now $K_l$ is the total electromagnetic field due to the seed plus the stimulated radiation, which is a dynamic variable evolving throughout the interaction and depending on the entire history of the e-beam spot size, current profile, and change in resonant energy throughout the interaction. The result is a complicated delay differential equation for $K_l$ where three-dimensional effects play an important role. In this case, it is easier to optimize the undulator tapering by solving for the actual field evolution with the help of 3D simulations.

In practice, the well-benchmarked 3D FEL simulation code Genesis \cite{Reiche:Genesis} is used to solve for the intensity of the radiation sampled by the electrons after a small number of undulator periods without tapering. This value is then fed into the tapering equation to calculate the optimum change of parameters for the following undulator section. These parameters along with the recorded particle and radiation distributions from the previous simulation are read by Genesis for the next step of the calculation which evolves the system for another small section of undulator. The optimization algorithm is then repeated until the end of the undulator. The result is an optimum undulator tapering and a self-consistent 3D simulation of the evolution of the electron beam and radiation in the optimized tapered undulator.

This tapering generation algorithm, dubbed Genesis informed tapering or GIT, can be used both in the constant and the varying period cases. For the former, Eqn. \ref{Equation:KTaper} is used to calculate the undulator $K$ parameter variation while the period is held constant to the initially assigned value. For the variable period undulator, Eqn. \ref{Equation:PeriodTaper} is used where $K$ is related to the undulator period by the undulator builder equation.

It is of critical importance to choose the variation of the undulator parameters in order to maintain the majority of the particles trapped in the ponderomotive bucket. In particular due to 3D effects, not all particles experience the same laser intensity or $K_l$. In order to account for this problem, GIT can look up the local intensity seen by each macro-particle in the simulation and soften the tapering to keep any desired fraction of the beam trapped within the resonant bucket.

%\section{Three-dimensional simulations}

\begin{figure}
\includegraphics[width=85mm]{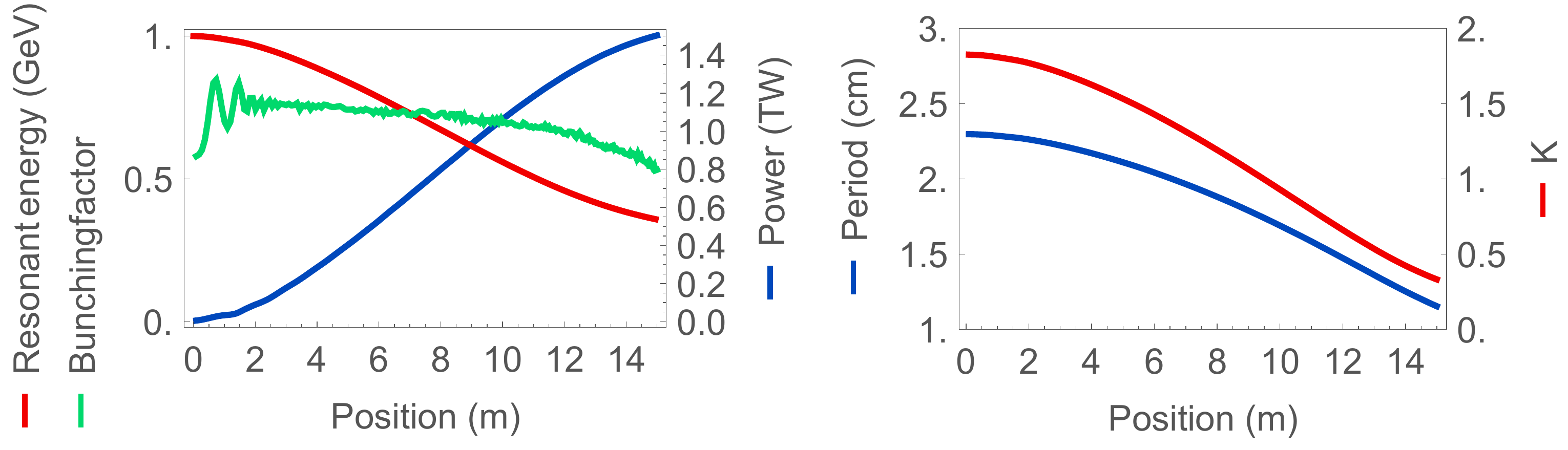}
\caption{The results of a GIT optimization for a 1~GeV, 3~kA beam lasing at 13.5~nm are shown in the above plots. The left plot shows the radiation power, resonant energy, and bunching factor versus position along the undulator while the right plot shows the undulator period and strength $K$ for the GIT optimized tapering.}
\label{Figure:EUV}
\end{figure}

\begin{figure*}
%\begin{figure*}[H]
\includegraphics*[width=160 mm]{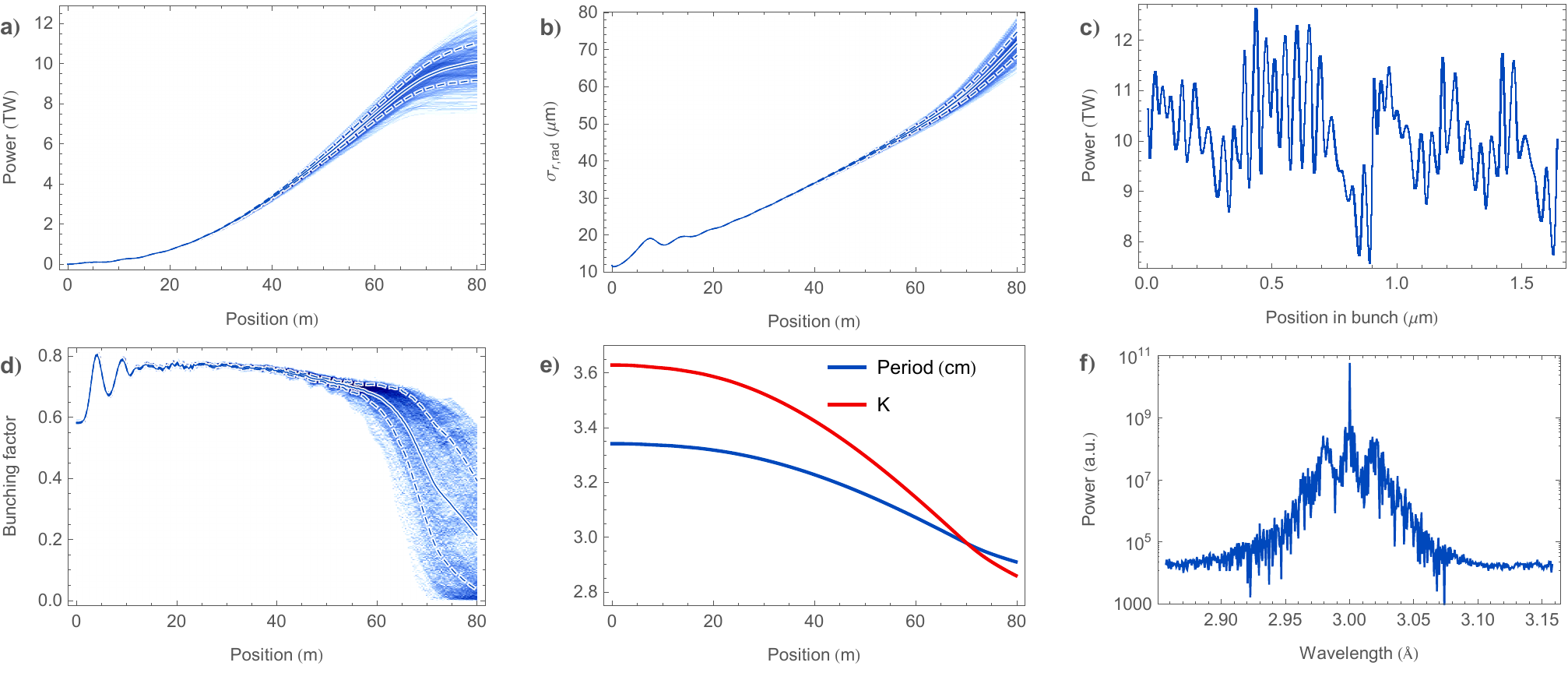}
\caption{GIT simulations for LCLS parameters. Shown along the undulator are the a) power, b) radiation size, d) bunching factor, and e) undulator period and $K$. The mean (solid line) as well as one standard deviation from the mean (dashed lines) of the values of all slices along each point in the undulator are also shown overlaid upon density plots of those values. The time-resolved power and spectrum at the undulator's exit is shown in figures c) and f) respectively.}
\label{Figure:TimeDependentLCLS}
\end{figure*}

In the rest of this paper we examine the results of GIT for a couple of relevant cases where high extraction efficiency can enable break-through applications for electron beam-based light sources. In the first example we consider a 1~GeV linac-driven radiation source for EUV lithography, which requires achieving high average power at 13.5~nm. In a conventional SASE FEL optimized for this wavelength range, a practically achievable Pierce parameter is on the order of $\sim$0.002, thus a state-of-the-art superconducting RF light source, such as XFEL, with about ~50 $\mu$A average current, can achieve about 100~W output. Applying a conventional adiabatic tapering technique could possibly increase the power output to 300-500~W range, which is still insufficient to meet industry needs. On the other hand, using a refocused EUV seed to drive a TESSA amplifier, it is possible to convert over 50\% of the electron beam power into the 13.5~nm light, all within a 15-m long TESSA undulator. Figure~2 shows the radiation power increase from 5~GW to $>$1.5~TW as the electron beam is decelerated in the process from 1~GeV to 320~MeV. Note that this remarkable numerical result still corresponds to a relatively modest decelerating gradient value of about 50~MeV/m, something that has already been demonstrated experimentally in the inverse (IFEL) configuration. Applying this result to the same example of a 50~$\mu$A XFEL-like driver beam with the TESSA afterburner one can achieve $>$20~kW power output at 13.5~nm---well within the application target range.

\begin{table}[hbt]
\centering
\caption{Simulation Parameters}
\label{Table:SimParam}
\setlength\tabcolsep{4pt}
\begin{ruledtabular}
\begin{tabular}{lcc}
\textbf{Parameter} & \textbf{EUV 13.5 nm} & \textbf{LCLS 3 \text{\AA}} \\
\hline
E-beam energy & 1 $\rightarrow$ 0.35\,GeV & 14.35 $\rightarrow$ 11.84 \,GeV \\
E-beam current & 3\,kA & 4\,kA \\
E-beam emittance & 0.5\ mm-mrad, & 0.2\,mm-mrad \\
E-beam spot size & 18\,$\mu$m & 9\,$\mu$m \\
Laser power & 5\,GW $\rightarrow$ 1.5\,TW & 5\,MW $\rightarrow$ 10.0\,TW \\
Seed Rayleigh range & 1\,m & 3\,m \\
Seed waist & 3\,m & 10\,m \\
Resonant phase & -1.00 $\rightarrow$ -0.78 & -1.57 $\rightarrow$ -1.50 \\
%Intensity percentile & 1st & 1st \\
Undulator period & 2.3 $\rightarrow$ 1.2 cm & 3.34 $\rightarrow$ 2.91 cm \\
Undulator K & 1.83 $\rightarrow$ 0.3 & 3.63 $\rightarrow$ 2.86 \\
Undulator length & 15 m & 80 m \\
\end{tabular}
\end{ruledtabular}
\end{table}

In the second example we consider the LCLS case where diffraction does not play a significant role due to the very short wavelength under study. The challenge here is to maximize the energy per pulse in order to enable single molecule imaging. More than 10$^{13}$ photons in a $<$10~fs pulse are required in order to beat the damage and obtain the diffraction information before destroying single molecules \cite{Chapman:diffractanddestroy}. For 4~keV photons (3~\text{\AA} wavelength), the peak power corresponding to this pulse approaches 1~TW.

We start our simulation from 5~MW power which is the typical level after self-seeding \cite{selfseeding}. More than 10~TW peak power levels can be obtained after 80~m of undulator as shown in Figure~\ref{Figure:TimeDependentLCLS}. For a 10~fs electron beam, this simulated output power corresponds to $\sim10^{14}$ photons per pulse even neglecting 3~fs slippage at the head and tail of the beam.

%\subsection{Time dependent simulations}

%\subsection{Sidebands}

An important effect is uncovered by the time-dependent simulations. When trapped in the ponderomotive potential, the electrons undergo synchrotron oscillations in longitudinal phase space and sideband frequencies are generated as pointed out in \cite{Wu:prstab}. In the time-domain, these correspond to oscillations in the time-profile of the field amplitude. This effect is clear in the simulation results shown in Figure~\ref{Figure:TimeDependentLCLS} where we follow a 6 fs slice of the beam along the 80 m long undulator. The increasing ripple in the temporal power profile appear in the spectrum as two sidebands around the central resonant frequency. As the amplitude of the oscillation grows, particles in those slices experiencing lower laser intensities detrap from the ponderomotive bucket, and the efficient energy exchange stops.

This synchrotron sideband instability is somewhat mitigated by strong tapering as the synchrotron frequency quickly changes along the interaction \cite{Sprangle:sidebands}. Furthermore, the GIT optimization algorithm can be performed for time-dependent simulations in the same way as we proceed for three dimensional effects, by making sure that the largest portion of the beam (now both transversely and longitudinally) can follow the tapering curve.

%\section{Conclusions}

In conclusion, TESSA is a novel approach to convert a significant fraction (50\% in one of our examples) of the energy of a relativistic electron beam into radiation by using tapering optimization techniques developed in the design of high gradient laser accelerators. This paper provides a physical basis for choosing the optimum tapering by dynamically matching the resonant energy gradient set by the undulator to the ponderomotive gradient due to the combined undulator and radiation fields.

The TESSA mechanism is ideal to take advantage of the relatively high wall-plug efficiency of particle accelerators. Maintaining high conversion efficiencies from wall to e-beam to radiation may allow the production of ultrahigh average power visible light sources with a wide range of applications including fusion science, defense and optically-driven accelerators for high luminosity colliders \cite{ESL}. In the near term, the single pass efficiency enhancement brought by TESSA can be used to generate coherent X-ray pulses of unprecedented intensity.

%\sections{Acknowledgments}

\acknowledgements{The authors are grateful for helpful discussions with Avi Gover, Claudio Emma, and Gerard Andonian.}

\bibliographystyle{unsrt}

\pagebreak

\end{document}